\documentstyle[multicol,aps,pra,epsfig,floats,amstex,here]{revtex}

\newcommand{\lv}{\left \vert}
\newcommand{\rv}{\right \vert}
\newcommand{\la}{\left \langle}
\newcommand{\ra}{\right \rangle}

\title{Remote information concentration\\
    using a bound entangled state}

\author{Mio Murao$^1$ and Vlatko Vedral$^2$}

\address {$^1$Semiconductors Laboratory,  The Institute of Physical
and Chemical Research (RIKEN), Wako-shi 351-0198, Japan \\
$^2$Centre for Quantum Computation,  Clarendon Laboratory,
University of Oxford, Oxford OX1 3PU, United Kingdom }

\date{\today}

\begin{document}
\draft
\maketitle

\begin{abstract}

Remote information concentration, the reverse process of quantum
telecloning, is presented. In this scheme, quantum information
originally from a single qubit, but now distributed into three
spatially separated qubits, is remotely concentrated back to a
single qubit via an initially shared entangled state without
performing any global operations. This entangled state is an
unlockable bound entangled state and we analyze its properties.
\end{abstract}


\begin{multicols}{2}

Quantum entanglement has generated a great deal of interest
recently as the most important resource in quantum information
processing. The protocols of super dense coding \cite{Bennett1},
quantum teleportation \cite{Teleportation} and telecloning
\cite{Telecloning} cannot be performed without some form of
entanglement between the parties involved: classical correlations
alone can never achieve the quantum efficiency arising from
entanglement. Given that entanglement is a resource, it is
important to be able to quantify it in order to deduce how
effectively we can process information. There have been a number
of suggestions for quantifying entanglement \cite{review}, but the
most fruitful method comes from a procedure known as entanglement
distillation \cite{Bennett2}. In this procedure, two distant
users, Alice and Bob, share a certain number of entangled pairs
all in the same state $\rho$. They then are allowed to perform
local operations and communicate classically with each other
(LOCC). The question is how many maximally entangled pairs they
can obtain in this way. The limit of distillation in the infinite
number of initial copies of $\rho$ is known as the entanglement of
distillation \cite{Bennett2}. A natural question to ask is: which states $\rho$
can be distilled to maximally entangled states? Separable states
$\rho=\sum_i \rho_A^i\otimes \rho_B^i$ are clearly
non-distillable. Surprisingly, however, a recent important
discovery by the Horodecki family showed that there are also some
entangled states which cannot be distilled \cite{Horodecki}. These
states have appropriately been called bound entangled. They are
peculiar as entanglement has to be invested in creating them by
LOCC, but this invested entanglement cannot then be recovered by
LOCC. Bound entanglement has been studied extensively in the last
two years \cite{Lewensteinreview}, nevertheless no information
processing protocol has been found where bound entangled states
perform better than just classically correlated states. Therefore
it has seemed that they are useless for quantum information
processing and that we always need to use some form of ``free''
(unbound) entanglement to achieve greater-than-classical
efficiency. However, as we show in this letter, this is not the
case.

We present an important protocol where bound entanglement can be
utilized effectively and performs better than any classically
correlated states. This protocol is remote information
concentration, the inverse of telecloning \cite{Telecloning}.
Quantum telecloning, as its name suggests, combines teleportation
and cloning in such a way that a sender teleports an unknown qubit
state $\lv \phi \ra=\alpha \lv 0 \ra + \beta \lv 1 \ra$ to a
number of spatially separated receivers simultaneously. These
teleported qubits cannot, of course, be exact replicas of the
original qubit due to the linear laws of quantum evolution
(``no-cloning theorem'') \cite{nocloning}. However it has been
shown that fidelities as high as allowed by the non-exact cloning
(known as optimal cloning \cite{Cloning}) can be achieved. The
optimal cloning state  for $\lv \phi \ra$ is represented by a
three qubit state
\begin{eqnarray}
    \lv \psi_c \ra &=&
    \alpha \sqrt{\frac{2}{3}}
    \left \{
    \lv 0 \ra \lv 00 \ra
    +
    \frac{1}{2} \lv 1 \ra
    \left (
    \lv 01 \ra + \lv 10 \ra
    \right )
    \right \} \nonumber \\
    &+&
    \beta
    \sqrt{\frac{2}{3}}
    \left \{
    \lv 1 \ra \lv 11 \ra
    +
    \frac{1}{2}
    \lv 0 \ra
    \left (
    \lv 01 \ra + \lv 10 \ra
    \right )
    \right
    \}.
\end{eqnarray}
where the first qubit is an ancilla and the last two qubits are
two optimal clones.  Now the question we ask is: once a state has
been telecloned to spatially separated parties, can it then be
recreated using only LOCC? The answer is yes and surprisingly
involves a recently constructed unlockable bound entangled state
\cite{Unlockable}.

The four particle unlockable bound entangled state presented by
Smolin \cite{Unlockable} is
\begin{eqnarray}
    \rho_{ub}=\frac{1}{4}
    \sum_{i=0}^{3}{
    \lv \Phi^i \ra \la \Phi^i \rv
    \otimes
    \lv \Phi^i \ra \la \Phi^i \rv
    }
\label{eqn:unlockablebestate}
\end{eqnarray}
where $\lv \Phi^i \ra$ represents the four Bell states, $\lv
\Phi^0 \ra = \left(\lv 00 \ra + \lv 11 \ra \right)/\sqrt{2}$, $\lv
\Phi^1 \ra = \left(\lv 00 \ra - \lv 11 \ra \right)/\sqrt{2}$, $\lv
\Phi^2 \ra = \left(\lv 01 \ra + \lv 10 \ra \right)/\sqrt{2}$ and
$\lv \Phi^3 \ra = \left(\lv 01 \ra - \lv 10 \ra \right)/\sqrt{2}$.
This state is not distillable if we do not allow joint quantum
operations (i.e. if all four parties only operate locally), and is
therefore, a bound entangle state.  However, if we allow a two
qubit joint operation, i.e. Bell joint measurement on any two
qubits, we can obtain a maximally entangled state for the other
two qubits via LOCC.  Thus this state is unlockable. The unlocking
mechanism is based on a joint operation for two out of four
qubits.

Before we explain remote information concentration, we briefly
summarize the forward process, telecloning. We focus on the $1$ to
$2$ telecloning and its reverse in this letter. Generalizations to
more qubits are possible and will be investigated elsewhere. The
telecloning scheme \cite{Telecloning} allows {\it direct}
distribution of optimal clones from an single original qubit state
$\lv \phi \ra$ to spatially separated parties using LOCC. In the
telecloning scheme, we use an initially shared entangled channel
(telecloning state)
\begin{eqnarray}
    \lv \xi_{tc} \ra
    &=&
    \frac{1}{\sqrt{3}} \left \{
    \lv 00 \ra \lv 00 \ra
    + \lv 11 \ra \lv 11 \ra
    \right.\\
    &+&\frac{1}{2} \left. \left( \lv 01 \ra
    + \lv 01 \ra \right )
    \left( \lv 01 \ra
    + \lv 01 \ra \right ) \right\}.
\end{eqnarray}
where the first qubit is an input port of the distributor, the
second qubit is an output port for the ancilla, and the third and
forth qubits are output ports for the optimal clones. The
telecloning protocol \cite{Telecloning} is similar to
teleportation; the distributor performs a Bell joint measurement
between the unknown state and the input port qubit, and then the
receivers, who hold output port qubits, perform a single qubit
operations depending on the distributor's measurement result.

\begin{figure}[H]
    \begin{center}
    \epsfig{file=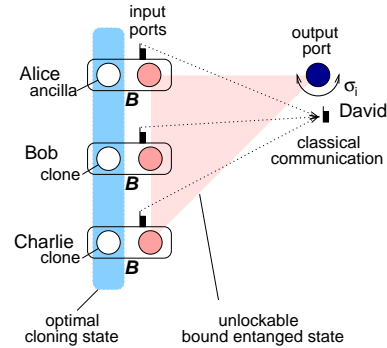,width=5cm}
    \end{center}
   \caption{Schematic picture showing the concentration of
   information from Alice, Bob and Charlie at the remote
   receiver, David, using an unlockable bound entangled
   state.}
    \label{fig:rci-ulbe}
\end{figure}

Now we present our remote information concentration scheme. From
the distributed optimal cloning qubits shared by the spatially
separated parties (Alice who holds the ancilla qubit and Bob and
Charlie who each hold a clone qubit), the original single qubit
state is recreated at the location of an receiver, David in our
scheme: $\lv \psi_c \ra_{ABC} \rightarrow \lv \phi \ra_D$.  We
employ the unlockable bound entangled state
(Eq.(\ref{eqn:unlockablebestate})) as an entangled channel for
this scheme.  The four qubits of the unlockable bound entangled
state are initially distributed to Alice, Bob and Charlie (input
port qubits) and David (output port qubit).  The three senders,
Alice, Bob and Charlie, {\it individually} perform Bell joint
measurements between their qubits of the optimal cloning state and
their input port qubits. We stress that no global operation is
allowed between qubits belonging to different parties. One of the
four possible outcomes $\{ \Phi^i \}$ is obtained by the
measurement of a party.  All three senders classically communicate
their measurement results with David. ($2 \times 3=6$ bits of
classical information is communicated in total.)  Each Bell
measurement result $\{ \Phi^i \}$ is associated with the
corresponding Pauli operators $\{ {\mathbf \sigma}_i \}$, where
${\mathbf \sigma}_0 \equiv {\bf 1}$, ${\mathbf \sigma}_1 \equiv
{\mathbf \sigma}_z$, ${\mathbf \sigma}_2 \equiv {\mathbf
\sigma}_x$, and ${\mathbf \sigma}_3 \equiv {\mathbf \sigma}_z
\cdot {\mathbf \sigma}_x$. David performs a Pauli operation ${\bf
\sigma}_j$, which is the product (up to a global phase factor) of
the three Pauli operators associated with the three Bell
measurements, on his output port qubit.  The output port of David
is now in the original state $\lv \phi \ra$. A schematic picture
of this protocol is shown in Fig.\ref{fig:rci-ulbe}. Since we do
not allow joint operations on spatially separated qubits, the
information channel in our scheme is indeed bound entangled. It is
surprising that a bound entangled state can actually be useful for
``transmitting'' quantum information.  In the following, we
analyze this feature from two points of view: remote quantum
operation and entanglement structure.

Remote quantum operation is performance of (global) unitary
operations on remote qubits:  A unitary operation $U$ is
implemented by an initially shared entangled channel, Bell
measurement, classical communication and (simple) single qubit
operations, instead of directly running a quantum circuit. This is
a generalization of quantum teleportation \cite{Teleportation}.
Telecloning \cite{Telecloning} and quantum information
distribution via entanglement \cite{QID} are examples of remote
quantum operation for $1 \rightarrow N$ quantum optimal cloning
with $d$-level particles, which requires one input port and $2N-1$
output ports.  More general cases requiring more than one input
port have been studied by Gottesman and Chuang \cite{Qsoftware} in
the context of ``quantum computation using teleportation'' and
``quantum software''. The initially shared entanglement in the
remote quantum operation scheme functions as quantum software.
According to their result, unitary operations which belong to the
Clifford group \cite{Clifford} can be implemented remotely, if we
restrict the single qubit operations to be the Pauli operations.
The compounding qubits of the shared entanglement need not be in
the same location.  In this case, the share entanglement functions
as a transmission channel as well as quantum software. We consider
this most restricted case of all-separated qubits.

To implement a unitary operation $U$ on a state of three input
qubits ($\lv \psi \ra$), the entangled channel consists of three
input port qubits and three output port qubits. For unitary
operations that can be decomposed into {\sf CNOT} (controlled {\sf
NOT}) and Hadamard gates, which are members of the Clifford group
\cite{Clifford}, the entangled channel state is given by
\begin{eqnarray} \label{eqn:generalchannel}
    \lv \xi \ra =
    \sum_{\tilde{k}=\tilde{0}}^{\tilde{7}}
    {\lv \tilde{k} \ra  \otimes U
    \lv \tilde{k} \ra},
\end{eqnarray}
where $\tilde{k}$ is a 3-bit binary number, for example,
$\tilde{0} = 000$, $\tilde{1} = 001$, ..., $\tilde{7} = 111$. The
first three qubits are the input ports and the last three qubits
are the output ports.  All the qubits of this channel are
spatially separated from each other.  We assume that Alice, Bob,
Charlie, David, Elizabeth, Fred each hold one qubit of the channel
(in this order).  Alice, Bob and Charlie {\it individually}
perform Bell joint measurements on their input qubits (in the
state $\lv \psi \ra_{ABC}$, the qubits to be processed) and the
input port qubits.  David, Elizabeth and Fred perform an
appropriate Pauli operation depending on the measurement results
of Alice, Bob and Charlie. The mapping between measurement results
and Pauli operations is initially agreed. The final state of
David, Elizabeth and Fred is $U \lv \psi \ra_{DEF}$.

\begin{figure}[H]
    \begin{center}
    \epsfig{file=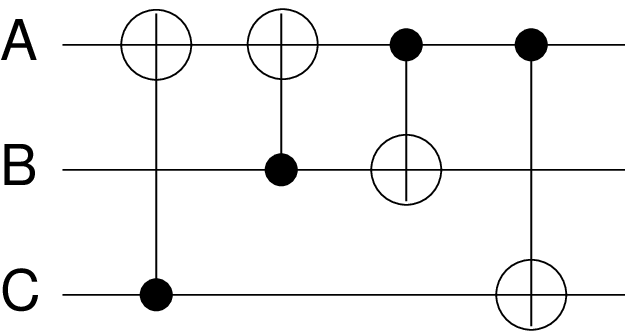,width=3cm}
    \end{center}
   \caption{Quantum circuit for the unitary operation of reverse
   cloning $U_r$.
   ``A'' represents an ancilla qubit, and ``B'' and ``C'' are the
   optimal clone qubits. The symbol ``$\bullet$'' represents the control
   qubit and they symbol ``$\oplus$'' represents the {\sf NOT} operation}
    \label{fig:rcnetwork}
\end{figure}

Now we return to reverse optimal cloning. We define a reverse
cloning unitary operator $U_r$
\begin{eqnarray}
    U_r \lv \psi_c \ra = \lv \phi \ra \otimes
    \sqrt{\frac{2}{3}}
    \left (
    \lv 00 \ra +
    \frac{\lv 01 \ra + \lv 10 \ra}{2}
    \right ),
\label{eqn:ureverseclone}
\end{eqnarray}
where the last two qubits are ancillas that are disentangled from
the first qubit which holds the concentrated single qubit
information.  Note that $U_r$ does not initialize the ancilla
qubits after the operation into the conventional ancilla state
$\lv 00 \ra$. $U_r$ can be decomposed into just CNOT gates as
shown in Fig.\ref{fig:rcnetwork}. Thus the reverse cloning
operation is in the Clifford group and can be performed by remote
quantum operation. $U_r$ is explicitly given in the computational
basis by
\begin{eqnarray}
    U_r=
    \begin{pmatrix}
      1 & 0 & 0 & 0 & 0 & 0 & 0 & 0 \\
      0 & 0 & 0 & 0 & 0 & 1 & 0 & 0 \\
      0 & 0 & 0 & 0 & 0 & 0 & 1 & 0 \\
      0 & 0 & 0 & 1 & 0 & 0 & 0 & 0 \\
      0 & 0 & 0 & 0 & 0 & 0 & 0 & 1 \\
      0 & 0 & 1 & 0 & 0 & 0 & 0 & 0 \\
      0 & 1 & 0 & 0 & 0 & 0 & 0 & 0 \\
      0 & 0 & 0 & 0 & 1 & 0 & 0 & 0 \
    \end{pmatrix}.
\end{eqnarray}
Inserting $U_r$ in Eq.(\ref{eqn:generalchannel}), we obtain the
channel for the remote reverse cloning:
\begin{eqnarray}
    \lv \xi_{rc} \ra
    &=& \frac{1}{2\sqrt{2}}
    \left \{
    \left ( \lv 0000 \ra + \lv 1111 \ra  \right )
    \lv 00 \ra
    +
    \left ( \lv 0101 \ra + \lv 1010 \ra  \right )
    \lv 01 \ra  \right.
    \nonumber \\
    &+&
    \left.
    \left ( \lv 0011 \ra + \lv 1100 \ra  \right )
    \lv 10 \ra
    +
    \left ( \lv 0110 \ra + \lv 1001 \ra  \right )
    \lv 11 \ra
    \right \}.
\label{eqn:rcchannel}
\end{eqnarray}
In this expression, Alice, Bob, Charlie and David hold the first,
second, third and fourth qubits, respectively.  The last two
qubits are ancillas and may be separated from the other qubits
(their location is irrelevant).  From Eq.(\ref{eqn:rcchannel}), we
obtain the unlockable bound entangled state $\rho_{ub}$, if we
trace out the ancilla variables.  Since the operations performed
on ancillas do not effect the output port qubit, we can trace out
the ancilla variables from the beginning. Then this remote quantum
operation is equivalent to remote information concentration. This
is why the unlockable bound entangled state actually functions as
a channel for remote information concentration.

Next, we analyze the remote information concentration scheme from
the view point of entanglement structure by considering the amount
of entanglement in various separation cuts. First, we consider the
pure state representation of the channel including the ancillas
(Eq.(\ref{eqn:rcchannel})). For simplicity, we denote the six
qubits as A, B, C, D, E, F.  From Eq.(\ref{eqn:generalchannel}),
we see that the channel can be considered as a pair of maximally
entangled 8-level particles, if we separate DEF from ABC.  This
means that there are 3-ebits of entanglement across the cut
ABC:DEF. This is why the information of three input qubits can be
processed and faithfully transmitted via teleportation from the
three senders to the three receivers.  We now consider the
entanglement structure of the unlockable bound entangled state. In
our scheme, at least 1-ebit of entanglement is required across the
ABC:D cut for the faithful transmission of a single qubit quantum
information.  We calculate the relative entropy of entanglement
\cite{Vedral} across the ABC:D cut. Relative entropy of
entanglement for a mixed state is defined by $E_{RE} =\min_{\sigma
\in D} S \left( \rho \vert \vert \sigma \right)$, where $S \left(
\rho \vert \vert \sigma \right)={\rm Tr} \rho \log \rho - \rho
\log \sigma $ is the quantum relative entropy, and the minimum is
taken over $D$, the set of separable states. We can prove that
in our case $E_{RE}=1$. The amount of entanglement
across the ABC:D cut is indeed 1-ebit. However, as we have
described before, there is no distillable pairwise entanglement in
the unlockable bound entangled state, if no joint operations are
allowed for qubits in different locations. Therefore the 1-ebit of
entanglement across ABC:D does not explain the successful
concentration of information. How can the information be processed
and faithfully transmitted only by LOCC?

The answer lies in the optimal cloning state. The three qubits in
the optimal cloning state are actually entangled each other. The
Peres-Horodecki criterion \cite{Peres-Horodecki}, which is the
smallest eigenvalue of partially transposed reduced density matrix
of an ancilla qubit and a clone qubit $\rho_{12} \equiv {\rm tr}_3
\lv \psi_c \ra \la \psi_c \rv$, is $(1-\sqrt{17})/12 \sim -0.26$.
This negativity shows that each of two clone qubits is entangled
(although not maximally entangled) with the ancilla qubits and
this entanglement is not bound entanglement.  We may conjecture
how information processing and transmission have been achieved
using only a bound entangled state and LOCC in our scheme as
follows: the Bell joint measurement ``combines'' the optimal
cloning state (the input state) and the bound entangled state (the
channel state), which is initially ``closed'' for transmission.
The unbound entanglement of the input state provides quantum
correlation among the qubits of the bound entangled state.  The
quantum correlation ``opens'' the channel for transmitting
concentrated single qubit information from distributed in three
qubits of the input state. Entanglement of the input optimal
cloning state and the unlockable bound entangled channel state
function in a complementary fashion. This result explains the
importance of the ancilla qubit in the optimal cloning state,
since the ancilla qubit is necessary for holding entanglement.

Another interesting observation is that the unlockable bound
entangled state is also valid for remotely concentrating
information from the spatially separate 3-qubit error correction
state: $\lv \psi_e \ra_{ABC}= \alpha \lv 000 \ra + \beta \lv 111
\ra \rightarrow \lv \phi \ra_D$. The procedure is similar to the
case of optimal cloning. The only difference is a modification to
the mapping to Pauli operations. David performs an additional
${\mathbf \sigma}_2$ if the measurement results from Bob or
Charlie, but not both, belong to the set $\left \{ \lv \Phi^{0}
\ra, \lv \Phi^{1} \ra \right \}$.  In this case, we may again
consider that the (unbound) entanglement of the input state $\lv
\psi_e \ra$ ``opens'' the bound entangled channel for transmitting
concentrated single qubit information from the input state.  If we
consider the quantum state $\lv \phi \ra$ as a quantum key
\cite{quantumkey}, remote information concentration together with
information distribution \cite {QID} may allow more secure
distribution of the quantum key to David via spatially-separated,
branched repeaters Alice, Bob and Charlie.

Finally, we show that no classically correlated state can achieve
the same task (c.f. \cite{review}). In optimal cloning scheme, due
to the linearity of quantum transformations, mixed states can be
cloned as well as pure. The same of course holds for telecloning.
We consider the case when the qubit to be telecloned is maximally
entangled with another qubit of George. After telecloning the
qubit state into the qubits of Alice, Bob and Charlie, we perform
the reverse process and remotely concentrate information at the
location of David.  Consequently, the qubits of David and George
become entangled. If a shared state with only classical
correlation could perform this remote information concentration,
the procedure would create entanglement between David and George.
This, however, is not possible: entanglement cannot be increased
by LOCC. Therefore no classically correlated state can
perform remote information concentration.

In this letter we have presented remote information concentration,
the reverse process of quantum telecloning. It was shown that,
surprisingly, the state needed for this operation is a bound
entangled state. We have analyzed the remote information
concentration scheme from two points of view, considering remote
quantum operations and analyzing the entanglement structure of the
bound state and the input state. We have shown that the unlockable
bound entangled state is a reduced density matrix for the
entanglement channel of remote reverse cloning, if we trace out
the ancilla qubits of the output state. From our entanglement
structure analysis, we have found that the functions of the
entanglement of the optimal cloning state and the unlockable bound
entangled state are complementary. We have also shown that the
unlockable bound entangled state can be used for remotely
concentrating information from a distributed 3-qubit error
correction state, which may be useful for secure transmission of a
quantum key. Furthermore, we showed that no purely classically
correlated state can achieve this task. We hope that our work
would stimulate more research into the nature of entanglement and
its general usefulness in quantum information processing.

The authors are grateful to J. Watson for his help during the
preparation of this manuscript.  M.M. is supported by the Special
Postdoctoral Researchers Program of RIKEN, and the Grant-in-Aid
for Encouragement of Young Scientists (Grant No. 12740253) by
Japan Society of the Promotion of Science. V.V. gratefully
acknowledges funding by the European Union project EQUIP
(Contract No. IST-1999-11053) and Hewlett Packard. Parts of this
work were completed during the Banasque Center for Science program
``Quantum Information Processing'' (2000).

\end{multicols}

\end{document}